\documentclass[twocolumn]{aastex63}

\usepackage{verbatim}
\usepackage{comment}
\usepackage{natbib}

\def\be{\begin{equation}}
\def\ee{\end{equation}}
\newcommand\code[1]{\textsc{\MakeLowercase{#1}}}

\def\gsim{\lower.5ex\hbox{\gtsima}} 
\def\lsim{\lower.5ex\hbox{\ltsima}} 
\def\gtsima{$\; \buildrel > \over \sim \;$} 
\def\ltsima{$\; \buildrel < \over \sim \;$} \def\gsim{\lower.5ex\hbox{\gtsima}} 
\def\lsim{\lower.5ex\hbox{\ltsima}} 
\def\simgt{\lower.5ex\hbox{\gtsima}} 
\def\simlt{\lower.5ex\hbox{\ltsima}}

\def\msun{{\rm M}_{\odot}}

\def\msunyr{\msun {\rm yr}^{-1}}

\def\S*{$\Sigma_{\rm SFR}$}

\def\CII{\hbox{[C~$\scriptstyle\rm II $]~}}

\def\OIII{\hbox{[O~$\scriptstyle\rm III $]~}}
 
\def\HI{\hbox{H~$\scriptstyle\rm I\ $}} 
\def\HII{\hbox{H~$\scriptstyle\rm II\ $}}

\definecolor{apcolor}{HTML}{b3003b}
\definecolor{afcolor}{HTML}{800080}
\definecolor{lvcolor}{HTML}{DF7401}
\definecolor{mkcolor}{HTML}{01abdf} 
\definecolor{cbcolor}{HTML}{ff0000}
\definecolor{sccolor}{HTML}{cc5500} 
\definecolor{sgcolor}{HTML}{00cc7a}

\makeatletter
\def\@hex@@Hex#1%
 {\if a#1A\else \if b#1B\else \if c#1C\else \if d#1D\else
  \if e#1E\else \if f#1F\else #1\fi\fi\fi\fi\fi\fi \@hex@Hex}
\makeatother
\definecolor{afcolor}{HTML}{b3443c}

\definecolor{apcolor}{HTML}{b3003b}

\revised{\today}
\shorttitle{ALMA non-detection of super-early blue galaxies}
\shortauthors{M. Kohandel et al.}
\graphicspath{{./}{figures/}}

\begin{document}

\title{Interpreting ALMA non-detections of JWST super-early galaxies}

\correspondingauthor{Mahsa Kohandel}
\email{mahsa.kohandel@sns.it}

\author[0000-0003-1041-7865]{M. Kohandel}
\affil{Scuola Normale Superiore,  Piazza dei Cavalieri 7, 50126 Pisa, Italy}

\author[0000-0002-9400-7312]{A. Ferrara}
\affil{Scuola Normale Superiore,  Piazza dei Cavalieri 7, 50126 Pisa, Italy}
\author[0000-0002-7129-5761]{A. Pallottini}
\affil{Scuola Normale Superiore,  Piazza dei Cavalieri 7, 50126 Pisa, Italy}
\author[0000-0002-3258-3672]{L. Vallini}
\affil{Scuola Normale Superiore,  Piazza dei Cavalieri 7, 50126 Pisa, Italy}
\author[0000-0002-2906-2200]{L. Sommovigo}
\affil{Scuola Normale Superiore,  Piazza dei Cavalieri 7, 50126 Pisa, Italy}
\author[0000-0001-6316-1707]{F. Ziparo}
\affil{Scuola Normale Superiore,  Piazza dei Cavalieri 7, 50126 Pisa, Italy}

\begin{abstract}
Recent attempts to detect [OIII] 88$\mu$m emission from super-early ($z>10$) galaxy candidates observed by JWST have been unsuccessful. By using zoom-in simulations, we show that these galaxies are faint, and mostly fall below the local metal-poor $\rm [OIII]-SFR$ relation as a result of their low ionization parameter, $U_{\rm ion}\simlt 10^{-3}$. Such low $U_{\rm ion}$ values are found in galaxies that are in an early assembly stage, and whose stars are still embedded in high-density natal clouds. However, the most luminous galaxy in our sample ($\rm{log}[L_{\rm{[OIII]}}/L_\odot] = 8.4$, $U_{\rm ion} \approx 0.1$) could be detected by ALMA in only $2.8$ hrs.
\end{abstract}


\section{Introduction} \label{sec:intro}

Early observations by the \textit{James Webb Space Telescope} (JWST) have discovered several bright ($\sim 60$ at $M_{\rm UV}\sim -21$) galaxy candidates at unprecedentedly high redshifts \citep[$z>10$:][]{Santini22,Adams22, Donnan22, Naidu22, Finkelstein22, Castellano22, Atek22, Whitler22, Harikane22, Furtak22, Yan22, Topping22, Rodighiero22}. If confirmed, the large abundance of these super-early systems is a challenge for galaxy formation models \citep{Ferrara22a, Finkelstein22, Mason22, Boylan-Kolchin22}. 

So far, these galaxies have been identified photometrically. Spectroscopic follow-ups are necessary to confirm their detection. Such a task becomes tricky at early cosmic times because the most common emission lines used at lower redshifts shift out the observable bands or (Ly$\alpha$) are severely suppressed by resonant scattering with intergalactic \HI at $z>6$. Detecting non-resonant emission lines (e.g. H$\alpha$) is the alternative strategy adopted by JWST NIRSpec observations. 

While waiting for such data, attempts have been made to confirm the detection of $z>10$ galaxies by using the Atacama Large sub-/Millimeter Array (ALMA) to search for fine-structure far-infrared (FIR) cooling lines, such as \OIII~$88\,\mu$m and \CII~$158\,\mu\rm{m}$ \citep{Bakx22, Popping22, Kaasinen22, Yoon22, Fujimoto22}. However, these experiments have been unsuccessful so far.

Among these galaxy candidates, GHZ2 has been independently detected by several groups \citep{Castellano22, Naidu22, Donnan22, Harikane22} which have reported a photometric redshift $z=11.960-12.423$. Using re-calibrated JWST fluxes \citet{Bakx22} have constrained the star formation rate of GZH2 in the range $\rm{SFR}=20^{+15}_{-14}\msunyr$ and stellar mass $M_\star = 1.2^{+7.2}_{-0.4}\times 10^8\msun$. \citet{Bakx22, Popping22} used ALMA Band 6 Director's Discretionary Time (DDT) program data to search for \OIII 88$\mu$m line emission. Their analysis has only provided a $5\sigma$ upper limit of $\log({L_{\rm{OIII}}}/L_\odot) < 1.7\times 10^{8}$ \citep{Bakx22} for the integrated \OIII line luminosity. 

Another case is GHZ1, a galaxy candidate from the JWST ERS program, GLASS-JWST \citep{Treu22} with a photometric redshift $z\sim10.6$, $\rm SFR = 36.3^{+54.5}_{-26.8}\msunyr$, and $\log(M_\star/\msun)=9.1^{+0.3}_{-0.4}$ \citep{Santini22}. The ALMA search for \OIII~$88\mu$m and dust continuum emission from GHZ1 was unsuccessful, albeit a marginal spectral feature was reported within $0.17^{''}$ of the JWST position of GHZ1 \citep{Yoon22}. The reported $5\sigma$ upper-limit of \OIII~luminosity is $2.2\times 10^8\,L_\odot$.

For a third candidate (HD1) a tentative $z=13.27$ was reported \citep{Harikane22a}. To confirm this detection, ALMA Band 6 (Band 4) observations were designed to target \OIII~$88\mu$m (\CII~158$\mu$m) emission \citep{Harikane22a, Kaasinen22}. No clear detection was found in either of the two BANDS. The $4\sigma$ upper limit for the \OIII~$88\mu$m  and \CII~158$\mu$m emission lines are $5.88\times 10^8$ and $0.7\times 10^8\,L_\odot$, respectively \citep{Kaasinen22}. 

The last one is S5-z17-1 identified in JWST ERO data of Stephan’s Quintet \citep{Pontoppidan22}, which has been followed-up by ALMA Band 7 \citep{Fujimoto22}, resulting in  $5.1\sigma$ line detection at 338.726 GHz, possibly corresponding to \OIII$52\mu\rm{m}$ at $z=16$. For this galaxy,  $\rm SFR <120\,\msunyr$.

These non-detections for $z>10$ sources discovered by JWST imply that either these galaxies are at lower redshifts ($z<4$), or they are too faint to be detected with the sensitivity of the above emission line experiments  \citep{Bakx22, Kaasinen22}. 

Thus, understanding the physical nature of the possible redshift evolution of the $L_{\rm [OIII]} - \rm SFR$ relation is of utmost importance in these early stages of interpretation of JWST data.
With this question in mind, here we analyse the \OIII$88\mu$m emission line from $z>11$ simulated galaxies.

\section{Simulated super-early galaxies}\label{sec:serra}

\code{SERRA} is a suite of zoom-in simulations that is tailored for EoR galaxies ($z \ge 6$; \citealt{Pallottini22}). In each radiation hydrodynamic simulation \citep{teyssier:2002,rosdahl:2013}, the comoving volume is $(20\,{\rm Mpc}/{\rm h})^{3}$ and contains $\sim 10-20$ galaxies, whose ISM is resolved on scales of $\simeq 1.2\times 10^4 \msun$ ($\simeq 30\, \rm pc$ at $z\sim 6$), i.e., the typical mass/size of molecular clouds \citep[e.g.][]{Federrath13}.

The non-equilibrium chemical network used in \code{SERRA} includes $\rm{H}$, $\rm{He}$, $\rm{H}^+$, $\rm{H}^-$, $\rm{He}$, $\rm{He}^+$, $\rm{He}^{++}$, $\rm{H}_2$, $\rm{H}_2^{+}$, and $e^-$ \citep{grassi:2014,pallottini:2017_b}. Metals are produced by stars \citep{pallottini:2017_a}, with solar relative abundances \citep{Asplund09}.
Line luminosities for each gas cell are computed in post-processing by using the spectral synthesis code \code{CLOUDY} \citep{Ferland17}, accounting for the turbulent and clumpy structure of the ISM \citep{Vallini17, Vallini18}, computed self-consistently from the simulation \citep{pallottini:2019}.
Given a field of view and a line of sight direction, the simulated galaxies can be mapped into 3-dimensional synthetic hyperspectral data cubes \citep{Kohandel20}, that can be directly compared with observations \citep{Zanella21, Rizzo22}.

In \code{SERRA}, we have 366 galaxies at $11\lsim z \lsim 14$ with $M_\star \simgt 10^8\msun$. Here, we extract a sub-sample with $6 \le \rm{SFR}/\msunyr \le 35$, matching the values measured for super-early JWST candidates; the final sample includes 42 galaxies. These are found in dark matter halos of mass $10^{10-10.7}M_\odot$, and are classified as \textit{starburst} based on their position on the $\Sigma_{\rm{SFR}}-\Sigma_{\rm{gas}}$ plane\footnote{The burstiness parameter $\kappa_s = 10^{12} \Sigma_{\rm{SFR}}/\Sigma_{\rm{g}}^{1.4}$ is defined in eq. 39 of \citet{Ferrara19}.} with $\kappa_s \sim 2-150$ \citep[][]{pallottini:2019}.

\section{[OIII] emission}\label{sec:FIR-serra}

\begin{figure}
  \centering\includegraphics[width = 0.5 \textwidth]{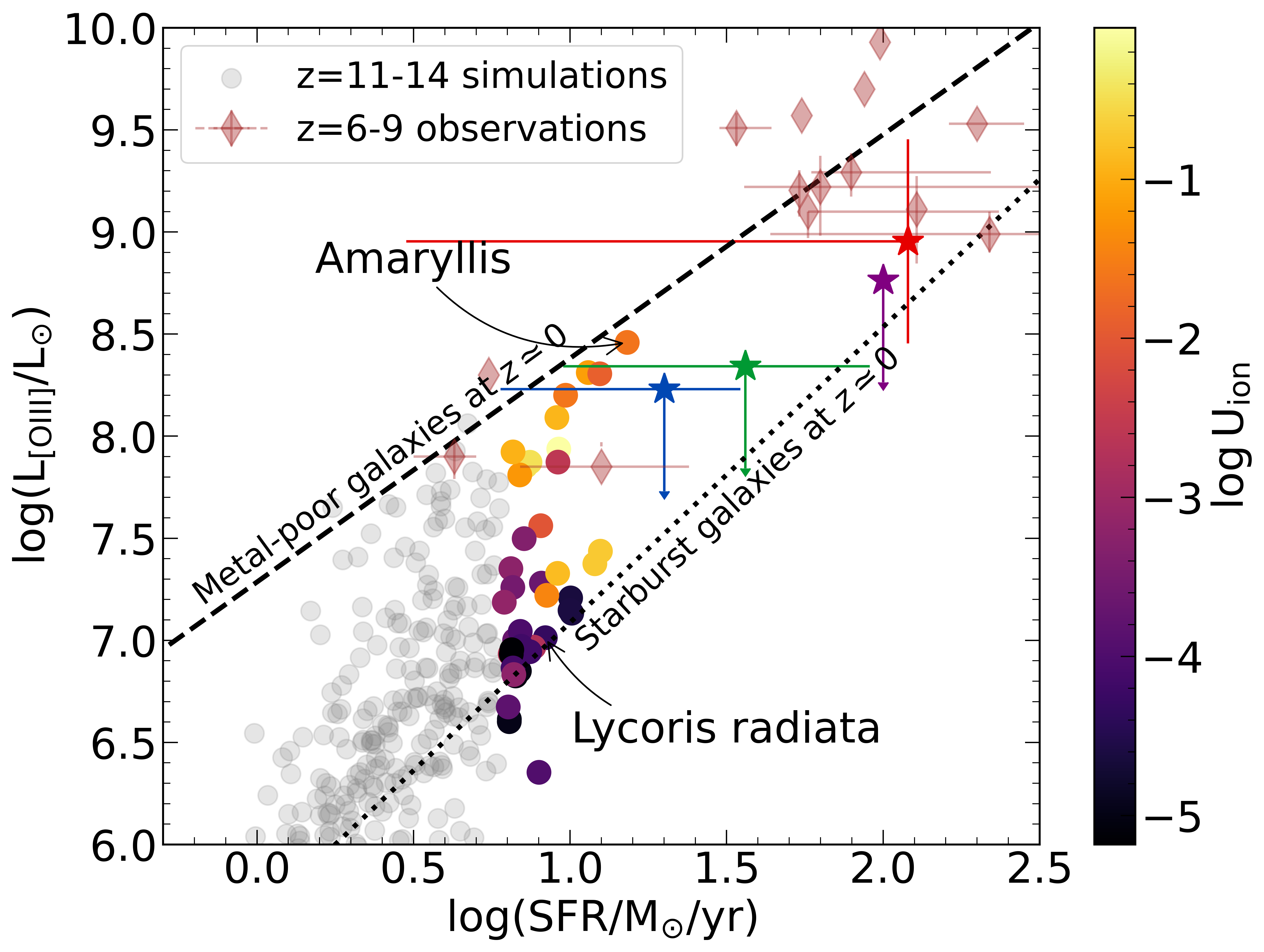}
  \caption{$L_{\OIII}-{\rm SFR}$ relation at redshift $11\lsim z \lsim 14$. Circles, colored according to the ionisation parameter $U_{\rm ion}$, represent \code{SERRA} galaxies with ${\rm SFR}> 6\, \msunyr$.
  Blue, green, violet and red stars indicate ALMA upper limits for JWST detected super-early galaxies: GHZ2 \citep{Bakx22}, GHZ1 \citep{Yoon22}, HD1 \citep{Kaasinen22}, and S5-z17-1 \citep{Fujimoto22}, respectively. Data from $z=6-9$ observations \citep[][diamonds]{Harikane20,Witstok22} are shown for comparison.  The $z=0$ relations for metal-poor and starburst galaxies \citep{deLooze14} are shown by dashed and dotted lines, respectively.
  \label{fig:Loiii-SFR-Uion}
  }
\end{figure}

Fig.~\ref{fig:Loiii-SFR-Uion} shows the position of \code{SERRA} galaxies at $z=11-14$ on the $L_{\rm{[OIII]}}-\rm{SFR}$ plane, along with the super-early JWST observed galaxies: GHZ2  \citep{Bakx22}, GHZ1 \citep{Yoon22}, HD1 \citep{Kaasinen22}, and S5-z17-1\footnote{Following \citet{Fujimoto22}, assuming that the emission feature is confirmed to be \OIII$52\mu\rm{m}$ line at $z=16$, it can be converted to \OIII$88\mu\rm{m}$ by assuming a conversion factor of 5.} \citep{Fujimoto22}. For comparison, we show $z=6-9$ galaxies from the literature \citep{Harikane20, Witstok22}, as well as the empirical relation for local metal-poor and starburst galaxies \citep{deLooze14}. 

\code{serra} galaxies generally fall between these two local populations, i.e., they are fainter than expected from the metal-poor relation. Moreover, although these galaxies have been selected to be in the relatively narrow SFR range $6-35\, \msunyr$, their $L_{\rm [OIII]}$ spans more than two orders of magnitude ($2.2-280 \times10^{6} L_\odot$). As a result, the local relation is effectively blurred as we move to the highest redshifts. This trend is also confirmed by the \code{SERRA} galaxies outside the JWST candidates SFR range (grey points) which are significantly fainter than expected from the local metal-poor relation. 

We note that the predicted luminosity of 3 \code{SERRA} galaxies is above the GHZ2 upper limit; hence it could have been detected. However, the detection probability critically depends on the width of the line.  The FWHM of the \OIII~line (face-on view\footnote{Face-on view corresponds to the narrowest emission line width \citep{Kohandel19}; thus, the estimated integration time below should be regarded as a lower limit.}) is in the range $107-325\,\rm{km~s}^{-1}$ with a mean of $213\,\rm{km~s}^{-1}$. If we consider the most luminous galaxy in our sample ($\rm{log}(L_{\rm{[OIII]}}/L_\odot) = 8.4$,   FWHM$=130\,\rm{km~s}^{-1}$) located at $z=12$, the total integration time required to detect it with ALMA at S/N$=5$ over the FWHM is $2.8$ hrs on-source, which is slightly longer than the integration times ($\approx 2$ hrs/tuning) of all ALMA DDT programs.

\section{Interpretation}\label{Sec:Inter}

\begin{figure}
\centering\includegraphics[width =  0.4 \textwidth]{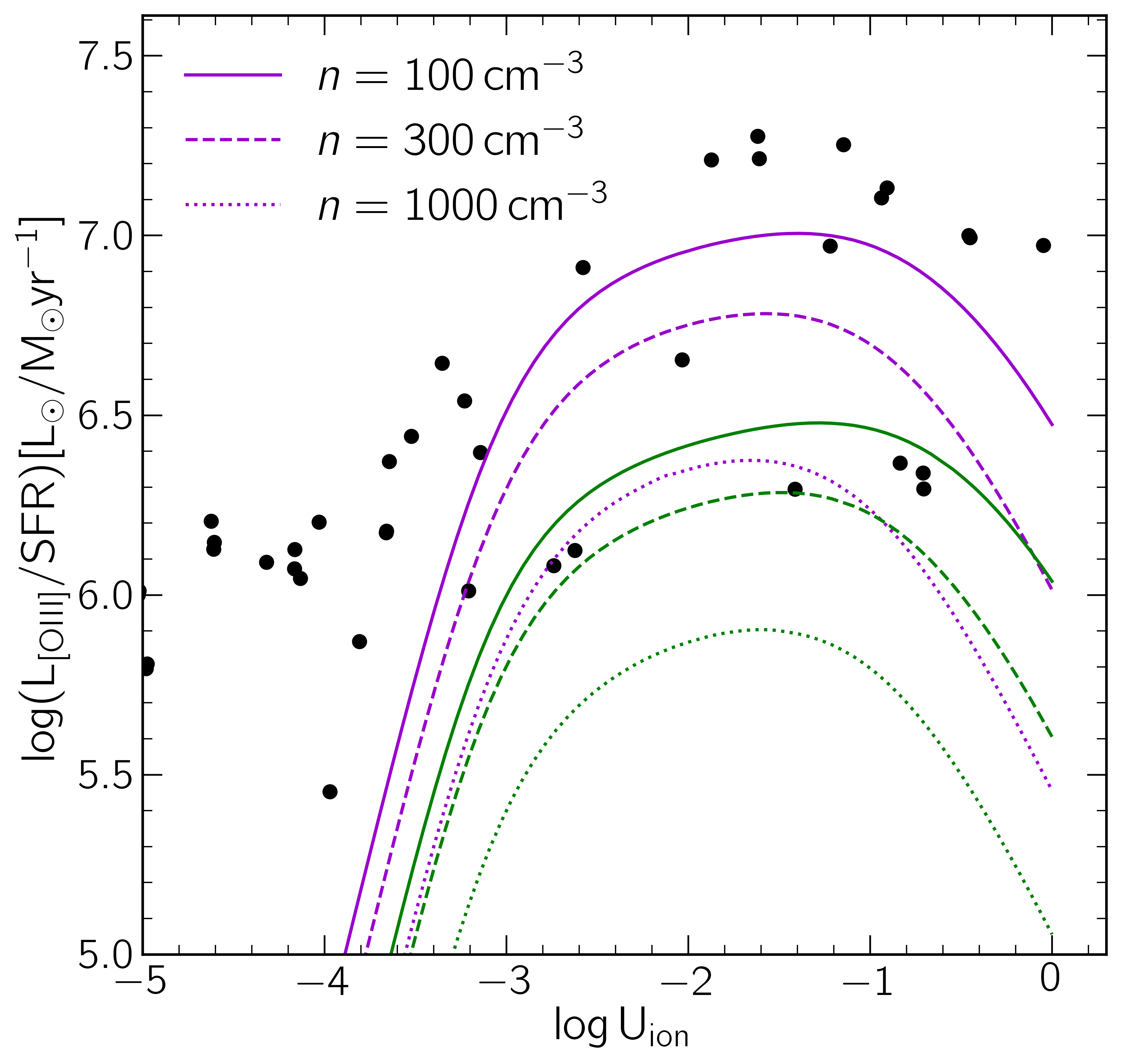}
\caption{$L_{\rm OIII}/\rm{SFR}$ ratio as a function of $U_{\rm ion}$ for \code{SERRA} galaxies ($z=11-14$, black circles) along with the results from single-zone \code{CLOUDY} models (lines) for different gas density, $n$, as shown in the legend. The purple (green) lines correspond to $Z=0.2\,Z_{\odot}$ ($Z=0.05\,Z_{\odot}$).
\label{Fig:harikane_models}}
\end{figure}

\begin{figure*}
\centering\includegraphics[width =  \textwidth]{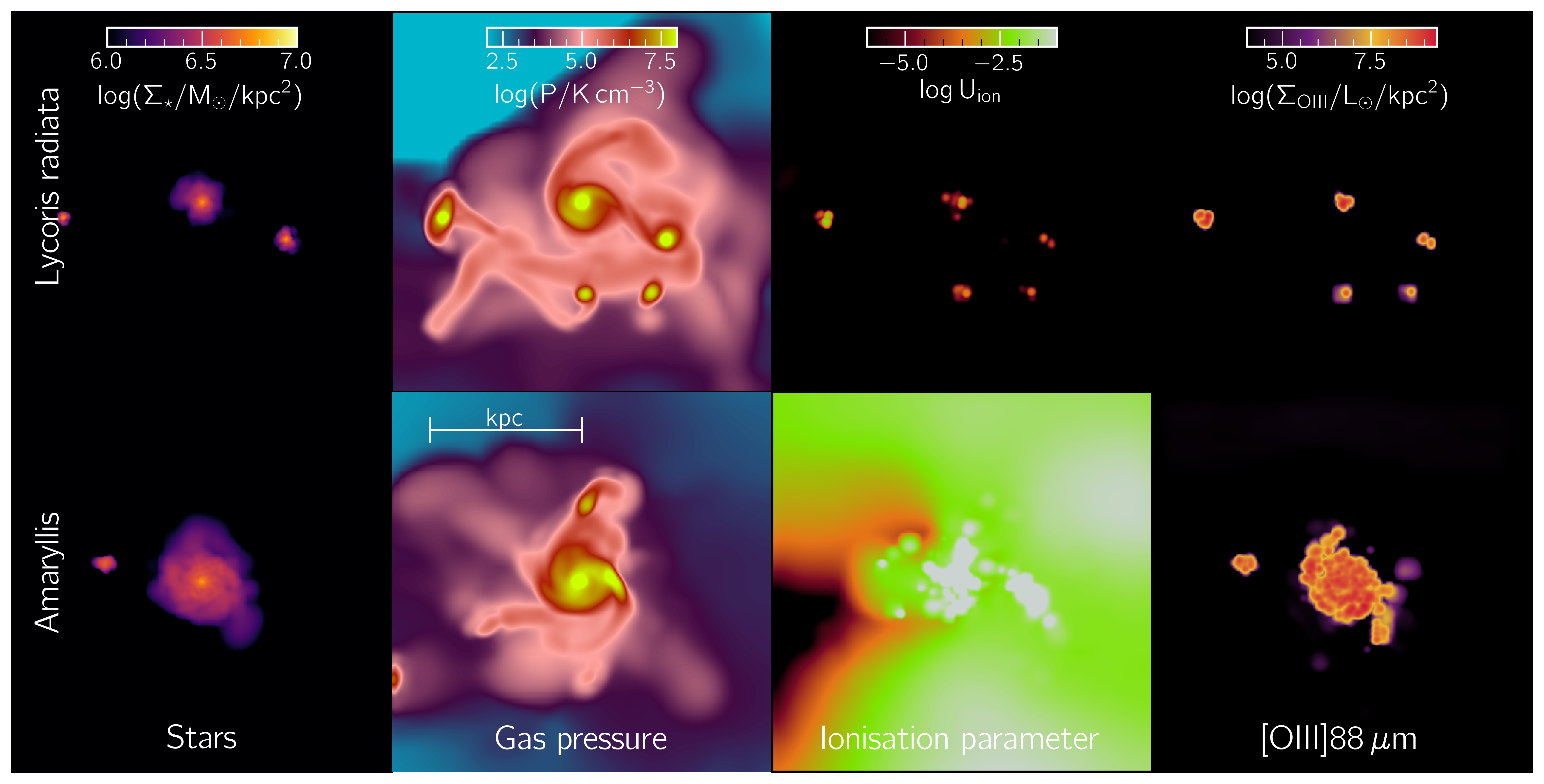}
\caption{Stellar, gas pressure, ionisation parameter, and \OIII~$88\mu$m emission maps for two \code{SERRA} galaxies with extremely low (\textit{Lycoris radiata}, upper panels) or high (\textit{Amaryllis}, bottom) ionisation parameters. Both systems are at $z=11.5$ and have similar SFR and $M_\star$ (see text). \textit{{Lycoris radiata}} is 20$\times$ fainter in \OIII~due to its very low $U_{\rm ion}$.
\label{Fig:lycoris}} 
\end{figure*}

Based on the evidence that \OIII~bright $z=6-9$ galaxies follow local metal-poor  $L_{\rm{[OIII]}}-\rm{SFR}$ relation \citep[][see points in Fig. \ref{fig:Loiii-SFR-Uion}]{Harikane20, Witstok22}, 4 different ALMA DDT experiments \citep{Bakx22,Treu22, Kaasinen22,Fujimoto22} have been designed accordingly to detect JWST $z>10$ candidates in \OIII~emission. However, all failed to detect the targeted sources at the expected luminosity, providing only upper limits.

\subsection{The role of the ionization parameter}

The first step to explain the unexpectedly low \OIII~luminosity of super-early galaxies is to consider single-zone \code{Cloudy} \citep{Ferland17} photoionisation models. Similarly to \citet{Harikane20}, we assume a plane-parallel gas slab with initial gas density $n$, metallicity $Z$, and illuminated by a radiation field with ionisation parameter\footnote{$U_{\rm ion} = n_\gamma/n$, where $n_\gamma$ is the density of $h\nu > 13.6$ eV photons.} $U_{\rm ion}$. The slab is set in pressure equilibrium ($P= 10^{4.5}n\, \rm K$); the computation is stopped at a depth $A_V=100$. The resulting $L_{\rm [OIII]}/\rm{SFR}$ ratio\footnote{The SFR is obtained from the emerging H$\alpha$ luminosity using the conversion factor in \citet[][eq. 2]{Kennicutt98}.} is plotted as a function of $U_{\rm ion}$ in Fig.~\ref{Fig:harikane_models} for representative models with $Z=(0.05, 0.2)\,Z_\odot$ and $n=10^{2-3}\, \rm{cm}^{-3}$. 

\code{Cloudy} models predict that weak \OIII~emission can be produced either by (i) low $Z$, (ii) high $n$, or (iii) low $U_{\rm ion}$.  All these quantities have an impact on \OIII~emission, but while $n$ and $Z$ show relatively small variations in the sample of simulated galaxies ($150 < n /\rm{cm}^{-3} < 1184$; $0.03 < Z/Z_\odot < 0.2$), the ionization parameter spans 5 orders of magnitude ($-5 \simlt \log U_{\rm{ion}} \simlt 0$). Hence, the strongest dependence of $L_{\rm [OIII]}$ is on $U_{\rm{ion}}$, in agreement with findings by \citet{Moriwaki18, Arata20}. The rapid drop of $L_{\rm [OIII]}$ for $\log U_{\rm{ion}} \simlt -3$ is only partially mitigated in the full RT, multi-phase \code{SERRA} results (black circles). In Fig.~\ref{fig:Loiii-SFR-Uion}, galaxies in our sample are color-coded with $U_{\rm ion}$. Indeed, galaxies with the largest downward deviations from the local metal-poor relation have $\log U_{\rm ion} < -3$. 

\subsection{What determines the ionization parameter?}

As galaxies in our sample, by construction, have very similar SFR values, what causes their wide ionization parameter range? Fig. \ref{Fig:lycoris} shows two galaxies from our sample: \textit{Lycoris radiata} and \textit{Amaryllis}. Shown are their stellar, gas pressure, ionisation parameter, and \OIII~$88\,\mu$m emission line maps. \textit{Lycoris radiata} is a galaxy with $M_\star = 6\times 10^{8}M_\odot$, ${\rm SFR}=10\,M_\odot \rm{yr}^{-1}$ at $z=11.5$ and \textit{Amaryllis} is at the same redshift with $M_\star = 6\times 10^{8}M_\odot$ and ${\rm SFR}=15\,M_\odot \rm{yr}^{-1}$. Although these two galaxies have similar SFR and are at the same redshift (Fig. \ref{fig:Loiii-SFR-Uion}), \textit{Amaryllis} is 20$\times$ brighter in \OIII due to its higher $U_{\rm ion}$, i.e. $7\times 10^{-2}$ vs. $5\times 10^{-5}$.

The difference in $U_{\rm ion}$ is mainly driven by the compactness of \HII regions. The ISM of early galaxies is highly pressurized as a result of the higher gas velocity dispersion. Fig.~\ref{Fig:lycoris} shows that the pressure within star-forming clumps is high, $P \approx 10^{7.5} \rm K\,cm^{-3}$, in both systems. As a consequence, clumps are more resistant to dispersal induced by stellar feedback, have longer lifetimes ($\simgt 10$ Myr), and star formation remains embedded for a longer time \citep{Behrens18, Sommovigo20}. These high-density ($n \approx 10^3 \, \rm cm^{-3}$) star-forming regions are characterized by $U_{\rm{ion}}\approx 10^{-5}$, which implies very low \OIII emission (Fig. \ref{Fig:harikane_models}). Also, ionising photons are trapped inside the clumps and produce a super-compact \HII region. 

While in \textit{Lycoris radiata} stars are almost exclusively located within high-density clumps, reflecting an earlier assembly stage in which stars are very young and still embedded, \textit{Amaryllis} has managed to build a well-developed disk structure, within which many star clusters have dispersed their natal cloud, ionised the low-density ISM, and produce a high $U_{\rm ion}$. This difference can be seen from the much more extended distribution of stars with respect to clumps in \textit{Amaryllis}. Thus, it is the ability to disperse natal clouds that ultimately determines the value of $U_{\rm ion}$ and the [OIII] luminosity.

\section{Summary}
To confirm the redshift of JWST detected super-early ($z>10$) galaxies, ALMA DDT observations have been designed aiming to detect the FIR \OIII~$88\,\mu$m emission line, supposing that these galaxies would follow the local metal-poor $L_{\rm{[OIII]}}-\rm{SFR}$ relation. Such follow-up observations have been unsuccessful so far. Here, we have used the \code{serra} suite of simulations of $z=11-14$ galaxies to explain such non-detections. We find that:

\begin{itemize}

\item[{\color{red} $\blacksquare$}]Galaxies with SFR similar to JWST-detected sources span more than two orders of magnitude in $L_{\rm{[OIII]}}$; most of them are fainter than expected from the $z=0$ metal-poor galaxies' relation. 

\item[{\color{red} $\blacksquare$}] The most luminous galaxy in our sample ($\rm{log}(L_{\rm{[OIII]}}/L_\odot) = 8.4$, FWHM$=130\,\rm{km~s}^{-1}$) could be detected in $2.8$ hrs on-source, which is slightly longer than the integration times ($\approx 2$ hrs/tuning) used by ALMA DDT programs so far.

\item[{\color{red} $\blacksquare$}] Galaxies with the largest downward deviations from the local metal-poor relation have low ionisation parameter, $\log U_{\rm ion} < -3$. 

\item[{\color{red} $\blacksquare$}] Such low $U_{\rm ion}$ values are found in galaxies that are in an early assembly stage, as most of their stars are still embedded in high-density natal clouds, and their ionizing photons are trapped in ultra-compact \HII regions.
\end{itemize}

\section*{Data Availability}
The derived data generated in this research will be shared on reasonable request to the corresponding author.

\acknowledgments
MK, AF, AP, LV, and LS acknowledge support from the ERC Advanced Grant INTERSTELLAR H2020/740120.
Generous support from the Carl Friedrich von Siemens-Forschungspreis der Alexander von Humboldt-Stiftung Research Award is kindly acknowledged (AF). 
We acknowledge the CINECA award under the ISCRA initiative, for the availability of high-performance computing resources and support from the Class B project SERRA HP10BPUZ8F.
We gratefully acknowledge computational resources of the Center for High Performance Computing (CHPC) at SNS.
We acknowledge usage of the Python programming language \citep{python2,python3}, Astropy \citep{astropy}, Cython \citep{cython}, Matplotlib \citep{matplotlib}, NumPy \citep{numpy}, \code{pynbody} \citep{pynbody}, and SciPy \citep{scipy}.

\bibliographystyle{aasjournal}
\bibliography{paper,codes}

\begin{thebibliography}{}
\expandafter\ifx\csname natexlab\endcsname\relax\def\natexlab#1{#1}\fi
\providecommand{\url}[1]{\href{#1}{#1}}
\providecommand{\dodoi}[1]{doi:~\href{http://doi.org/#1}{\nolinkurl{#1}}}
\providecommand{\doeprint}[1]{\href{http://ascl.net/#1}{\nolinkurl{http://ascl.net/#1}}}
\providecommand{\doarXiv}[1]{\href{https://arxiv.org/abs/#1}{\nolinkurl{https://arxiv.org/abs/#1}}}

\bibitem[{{Adams} {et~al.}(2022){Adams}, {Conselice}, {Ferreira}, {Austin},
  {Trussler}, {Juod{\v{z}}balis}, {Wilkins}, {Caruana}, \& {Dayal}}]{Adams22}
{Adams}, N.~J., {Conselice}, C.~J., {Ferreira}, L., {et~al.} 2022, arXiv
  e-prints, arXiv:2207.11217.
\newblock \doarXiv{2207.11217}

\bibitem[{{Arata} {et~al.}(2020){Arata}, {Yajima}, {Nagamine}, {Abe}, \&
  {Khochfar}}]{Arata20}
{Arata}, S., {Yajima}, H., {Nagamine}, K., {Abe}, M., \& {Khochfar}, S. 2020,
  \mnras, 498, 5541, \dodoi{10.1093/mnras/staa2809}

\bibitem[{Asplund {et~al.}(2009)Asplund, Grevesse, Sauval, \&
  Scott}]{Asplund09}
Asplund, M., Grevesse, N., Sauval, A.~J., \& Scott, P. 2009, Annual Review of
  Astronomy and Astrophysics, 47, 481–522,
  \dodoi{10.1146/annurev.astro.46.060407.145222}

\bibitem[{{Astropy Collaboration} {et~al.}(2013){Astropy Collaboration},
  {Robitaille}, {Tollerud}, {Greenfield}, {Droettboom}, {Bray}, {Aldcroft},
  {Davis}, {Ginsburg}, {Price-Whelan}, {Kerzendorf}, {Conley}, {Crighton},
  {Barbary}, {Muna}, {Ferguson}, {Grollier}, {Parikh}, {Nair}, {Unther},
  {Deil}, {Woillez}, {Conseil}, {Kramer}, {Turner}, {Singer}, {Fox}, {Weaver},
  {Zabalza}, {Edwards}, {Azalee Bostroem}, {Burke}, {Casey}, {Crawford},
  {Dencheva}, {Ely}, {Jenness}, {Labrie}, {Lim}, {Pierfederici}, {Pontzen},
  {Ptak}, {Refsdal}, {Servillat}, \& {Streicher}}]{astropy}
{Astropy Collaboration}, {Robitaille}, T.~P., {Tollerud}, E.~J., {et~al.} 2013,
  \aap, 558, A33, \dodoi{10.1051/0004-6361/201322068}

\bibitem[{Atek {et~al.}(2022)Atek, Shuntov, Furtak, Richard, Kneib, Zitrin, \&
  Charlot}]{Atek22}
Atek, H., Shuntov, M., Furtak, L.~J., {et~al.} 2022, Revealing Galaxy
  Candidates out to $z \sim 16$ with JWST Observations of the Lensing Cluster
  SMACS0723,  arXiv, \dodoi{10.48550/ARXIV.2207.12338}

\bibitem[{{Bakx} {et~al.}(2022){Bakx}, {Zavala}, {Mitsuhashi}, {Treu},
  {Fontana}, {Tadaki}, {Casey}, {Castellano}, {Glazebrook}, {Hagimoto},
  {Ikeda}, {Jones}, {Leethochawalit}, {Mason}, {Morishita}, {Nanayakkara},
  {Pentericci}, {Roberts-Borsani}, {Santini}, {Serjeant}, {Tamura}, {Trenti},
  \& {Vanzella}}]{Bakx22}
{Bakx}, T. J.~L.~C., {Zavala}, J.~A., {Mitsuhashi}, I., {et~al.} 2022, arXiv
  e-prints, arXiv:2208.13642.
\newblock \doarXiv{2208.13642}

\bibitem[{Behnel {et~al.}(2011)Behnel, Bradshaw, Citro, Dalcin, Seljebotn, \&
  Smith}]{cython}
Behnel, S., Bradshaw, R., Citro, C., {et~al.} 2011, Computing in Science
  Engineering, 13, 31 , \dodoi{10.1109/MCSE.2010.118}

\bibitem[{{Behrens} {et~al.}(2018){Behrens}, {Pallottini}, {Ferrara},
  {Gallerani}, \& {Vallini}}]{Behrens18}
{Behrens}, C., {Pallottini}, A., {Ferrara}, A., {Gallerani}, S., \& {Vallini},
  L. 2018, \mnras, 477, 552, \dodoi{10.1093/mnras/sty552}

\bibitem[{{Boylan-Kolchin}(2022)}]{Boylan-Kolchin22}
{Boylan-Kolchin}, M. 2022, arXiv e-prints, arXiv:2208.01611.
\newblock \doarXiv{2208.01611}

\bibitem[{{Castellano} {et~al.}(2022){Castellano}, {Fontana}, {Treu},
  {Santini}, {Merlin}, {Leethochawalit}, {Trenti}, {Mestric}, {Vanzella},
  {Bonchi}, {Belfiori}, {Nonino}, {Paris}, {Polenta}, {Roberts-Borsani},
  {Boyett}, {Calabro}, {Glazebrook}, {Grillo}, {Mascia}, {Mason}, {Mercurio},
  {Morishita}, {Nanayakkara}, {Pentericci}, {Rosati}, {Vulcani}, {Wang}, \&
  {Yang}}]{Castellano22}
{Castellano}, M., {Fontana}, A., {Treu}, T., {et~al.} 2022, arXiv e-prints,
  arXiv:2207.09436.
\newblock \doarXiv{2207.09436}

\bibitem[{{De Looze} {et~al.}(2014){De Looze}, {Cormier}, {Lebouteiller},
  {Madden}, {Baes}, {Bendo}, {Boquien}, {Boselli}, {Clements}, {Cortese},
  {Cooray}, {Galametz}, {Galliano}, {Graci{\'a}-Carpio}, {Isaak}, {Karczewski},
  {Parkin}, {Pellegrini}, {R{\'e}my-Ruyer}, {Spinoglio}, {Smith}, \&
  {Sturm}}]{deLooze14}
{De Looze}, I., {Cormier}, D., {Lebouteiller}, V., {et~al.} 2014, \aap, 568,
  A62, \dodoi{10.1051/0004-6361/201322489}

\bibitem[{Donnan {et~al.}(2022)Donnan, McLeod, Dunlop, McLure, Carnall, Begley,
  Cullen, Hamadouche, Bowler, McCracken, Milvang-Jensen, Moneti, \&
  Targett}]{Donnan22}
Donnan, C.~T., McLeod, D.~J., Dunlop, J.~S., {et~al.} 2022, The evolution of
  the galaxy UV luminosity function at redshifts z ~ 8-15 from deep JWST and
  ground-based near-infrared imaging,  arXiv, \dodoi{10.48550/ARXIV.2207.12356}

\bibitem[{{Federrath} \& {Klessen}(2013)}]{Federrath13}
{Federrath}, C., \& {Klessen}, R.~S. 2013, \apj, 763, 51,
  \dodoi{10.1088/0004-637X/763/1/51}

\bibitem[{{Ferland} {et~al.}(2017){Ferland}, {Chatzikos}, {Guzm{\'a}n},
  {Lykins}, {van Hoof}, {Williams}, {Abel}, {Badnell}, {Keenan}, {Porter}, \&
  {Stancil}}]{Ferland17}
{Ferland}, G.~J., {Chatzikos}, M., {Guzm{\'a}n}, F., {et~al.} 2017, \rmxaa, 53,
  385.
\newblock \doarXiv{1705.10877}

\bibitem[{{Ferrara} {et~al.}(2022){Ferrara}, {Pallottini}, \&
  {Dayal}}]{Ferrara22a}
{Ferrara}, A., {Pallottini}, A., \& {Dayal}, P. 2022, arXiv e-prints,
  arXiv:2208.00720, \dodoi{10.48550/ARXIV.2208.00720}

\bibitem[{{Ferrara} {et~al.}(2019){Ferrara}, {Vallini}, {Pallottini},
  {Gallerani}, {Carniani}, {Kohandel}, {Decataldo}, \& {Behrens}}]{Ferrara19}
{Ferrara}, A., {Vallini}, L., {Pallottini}, A., {et~al.} 2019, \mnras, 489, 1,
  \dodoi{10.1093/mnras/stz2031}

\bibitem[{{Finkelstein} {et~al.}(2022){Finkelstein}, {Bagley}, {Arrabal Haro},
  {Dickinson}, {Ferguson}, {Kartaltepe}, {Papovich}, {Burgarella}, {Kocevski},
  {Huertas-Company}, {Iyer}, {Larson}, {P{\'e}rez-Gonz{\'a}lez}, {Rose},
  {Tacchella}, {Wilkins}, {Chworowsky}, {Medrano}, {Morales}, {Somerville},
  {Yung}, {Fontana}, {Giavalisco}, {Grazian}, {Grogin}, {Kewley}, {Koekemoer},
  {Kirkpatrick}, {Kurczynski}, {Lotz}, {Pentericci}, {Pirzkal}, {Ravindranath},
  {Ryan}, {Trump}, {Yang}, {Almaini}, {Amor{\'\i}n}, {Annunziatella},
  {Backhaus}, {Barro}, {Behroozi}, {Bell}, {Bhatawdekar}, {Bisigello}, {Bromm},
  {Buat}, {Buitrago}, {Calabr{\'o}}, {Casey}, {Castellano}, {Ch{\'a}vez Ortiz},
  {Ciesla}, {Cleri}, {Cohen}, {Cole}, {Cooke}, {Cooper}, {Cooray}, {Costantin},
  {Cox}, {Croton}, {Daddi}, {Dav{\'e}}, {de la Vega}, {Dekel}, {Elbaz},
  {Estrada-Carpenter}, {Faber}, {Fern{\'a}ndez}, {Finkelstein}, {Freundlich},
  {Fujimoto}, {Garc{\'\i}a-Argum{\'a}nez}, {Gardner}, {Gawiser},
  {G{\'o}mez-Guijarro}, {Guo}, {Hamilton}, {Hathi}, {Holwerda}, {Hirschmann},
  {Hutchison}, {Jha}, {Jogee}, {Juneau}, {Jung}, {Kassin}, {Le Bail}, {Leung},
  {Lucas}, {Magnelli}, {Mantha}, {Matharu}, {McGrath}, {McIntosh}, {Merlin},
  {Mobasher}, {Newman}, {Nicholls}, {Pandya}, {Rafelski}, {Ronayne}, {Santini},
  {Seill{\'e}}, {Shah}, {Shen}, {Simons}, {Snyder}, {Stanway}, {Straughn},
  {Teplitz}, {Vanderhoof}, {Vega-Ferrero}, {Wang}, {Weiner}, {Willmer},
  {Wuyts}, \& {Zavala}}]{Finkelstein22}
{Finkelstein}, S.~L., {Bagley}, M.~B., {Arrabal Haro}, P., {et~al.} 2022, arXiv
  e-prints, arXiv:2207.12474.
\newblock \doarXiv{2207.12474}

\bibitem[{{Fujimoto} {et~al.}(2022){Fujimoto}, {Finkelstein}, {Burgarella},
  {Carilli}, {Buat}, {Casey}, {Ciesla}, {Tacchella}, {Zavala}, {Brammer},
  {Fudamoto}, {Ouchi}, {Valentino}, {Cooper}, {Dickinson}, {Franco},
  {Giavalisco}, {Hutchison}, {Kartaltepe}, {Koekemoer}, {Kojima}, {Larson},
  {Murphy}, {Papovich}, {P{\'e}rez-Gonz{\'a}lez}, {Somerville}, {Yoon},
  {Wilkins}, {Yung}, {Akins}, {Amor{\'\i}n}, {Arrabal Haro}, {Bagley},
  {Chworowsky}, {Cooper}, {Costantin}, {Daddi}, {Ferguson}, {Grogin},
  {Jim{\'e}nez-Andrade}, {Juneau}, {Kirkpatrick}, {Kocevski}, {Le Bail},
  {Long}, {Lucas}, {Magnelli}, {McKinney}, {Rose}, {Seill{\'e}}, {Simons}, \&
  {Weiner}}]{Fujimoto22}
{Fujimoto}, S., {Finkelstein}, S.~L., {Burgarella}, D., {et~al.} 2022, arXiv
  e-prints, arXiv:2211.03896.
\newblock \doarXiv{2211.03896}

\bibitem[{{Furtak} {et~al.}(2022){Furtak}, {Shuntov}, {Atek}, {Zitrin},
  {Richard}, {Lehnert}, \& {Chevallard}}]{Furtak22}
{Furtak}, L.~J., {Shuntov}, M., {Atek}, H., {et~al.} 2022, arXiv e-prints,
  arXiv:2208.05473.
\newblock \doarXiv{2208.05473}

\bibitem[{{Grassi} {et~al.}(2014){Grassi}, {Bovino}, {Schleicher}, {Prieto},
  {Seifried}, {Simoncini}, \& {Gianturco}}]{grassi:2014}
{Grassi}, T., {Bovino}, S., {Schleicher}, D. R.~G., {et~al.} 2014, \mnras, 439,
  2386, \dodoi{10.1093/mnras/stu114}

\bibitem[{{Harikane} {et~al.}(2020){Harikane}, {Ouchi}, {Inoue}, {Matsuoka},
  {Tamura}, {Bakx}, {Fujimoto}, {Moriwaki}, {Ono}, {Nagao}, {Tadaki}, {Kojima},
  {Shibuya}, {Egami}, {Ferrara}, {Gallerani}, {Hashimoto}, {Kohno}, {Matsuda},
  {Matsuo}, {Pallottini}, {Sugahara}, \& {Vallini}}]{Harikane20}
{Harikane}, Y., {Ouchi}, M., {Inoue}, A.~K., {et~al.} 2020, \apj, 896, 93,
  \dodoi{10.3847/1538-4357/ab94bd}

\bibitem[{{Harikane} {et~al.}(2022{\natexlab{a}}){Harikane}, {Ouchi}, {Oguri},
  {Ono}, {Nakajima}, {Isobe}, {Umeda}, {Mawatari}, \& {Zhang}}]{Harikane22}
{Harikane}, Y., {Ouchi}, M., {Oguri}, M., {et~al.} 2022{\natexlab{a}}, arXiv
  e-prints, arXiv:2208.01612.
\newblock \doarXiv{2208.01612}

\bibitem[{{Harikane} {et~al.}(2022{\natexlab{b}}){Harikane}, {Inoue},
  {Mawatari}, {Hashimoto}, {Yamanaka}, {Fudamoto}, {Matsuo}, {Tamura}, {Dayal},
  {Yung}, {Hutter}, {Pacucci}, {Sugahara}, \& {Koekemoer}}]{Harikane22a}
{Harikane}, Y., {Inoue}, A.~K., {Mawatari}, K., {et~al.} 2022{\natexlab{b}},
  \apj, 929, 1, \dodoi{10.3847/1538-4357/ac53a9}

\bibitem[{Hunter(2007)}]{matplotlib}
Hunter, J.~D. 2007, Computing in Science Engineering, 9, 90,
  \dodoi{10.1109/MCSE.2007.55}

\bibitem[{{Kaasinen} {et~al.}(2022){Kaasinen}, {van Marrewijk}, {Popping},
  {Ginolfi}, {Di Mascolo}, {Mroczkowski}, {Concas}, {Di Cesare}, {Killi}, \&
  {Langan}}]{Kaasinen22}
{Kaasinen}, M., {van Marrewijk}, J., {Popping}, G., {et~al.} 2022, arXiv
  e-prints, arXiv:2210.03754.
\newblock \doarXiv{2210.03754}

\bibitem[{{Kennicutt}(1998)}]{Kennicutt98}
{Kennicutt}, Robert~C., J. 1998, \araa, 36, 189,
  \dodoi{10.1146/annurev.astro.36.1.189}

\bibitem[{{Kohandel} {et~al.}(2020){Kohandel}, {Pallottini}, {Ferrara},
  {Carniani}, {Gallerani}, {Vallini}, {Zanella}, \& {Behrens}}]{Kohandel20}
{Kohandel}, M., {Pallottini}, A., {Ferrara}, A., {et~al.} 2020, \mnras, 499,
  1250, \dodoi{10.1093/mnras/staa2792}

\bibitem[{{Kohandel} {et~al.}(2019){Kohandel}, {Pallottini}, {Ferrara},
  {Zanella}, {Behrens}, {Carniani}, {Gallerani}, \& {Vallini}}]{Kohandel19}
---. 2019, \mnras, 487, 3007, \dodoi{10.1093/mnras/stz1486}

\bibitem[{{Mason} {et~al.}(2022){Mason}, {Trenti}, \& {Treu}}]{Mason22}
{Mason}, C.~A., {Trenti}, M., \& {Treu}, T. 2022, arXiv e-prints,
  arXiv:2207.14808.
\newblock \doarXiv{2207.14808}

\bibitem[{{Moriwaki} {et~al.}(2018){Moriwaki}, {Yoshida}, {Shimizu},
  {Harikane}, {Matsuda}, {Matsuo}, {Hashimoto}, {Inoue}, {Tamura}, \&
  {Nagao}}]{Moriwaki18}
{Moriwaki}, K., {Yoshida}, N., {Shimizu}, I., {et~al.} 2018, \mnras, 481, L84,
  \dodoi{10.1093/mnrasl/sly167}

\bibitem[{{Naidu} {et~al.}(2022){Naidu}, {Oesch}, {Dokkum}, {Nelson}, {Suess},
  {Brammer}, {Whitaker}, {Illingworth}, {Bouwens}, {Tacchella}, {Matthee},
  {Allen}, {Bezanson}, {Conroy}, {Labbe}, {Leja}, {Leonova}, {Magee}, {Price},
  {Setton}, {Strait}, {Stefanon}, {Toft}, {Weaver}, \& {Weibel}}]{Naidu22}
{Naidu}, R.~P., {Oesch}, P.~A., {Dokkum}, P.~v., {et~al.} 2022, \apjl, 940,
  L14, \dodoi{10.3847/2041-8213/ac9b22}

\bibitem[{{Pallottini} {et~al.}(2017{\natexlab{a}}){Pallottini}, {Ferrara},
  {Bovino}, {Vallini}, {Gallerani}, {Maiolino}, \&
  {Salvadori}}]{pallottini:2017_b}
{Pallottini}, A., {Ferrara}, A., {Bovino}, S., {et~al.} 2017{\natexlab{a}},
  \mnras, 471, 4128, \dodoi{10.1093/mnras/stx1792}

\bibitem[{{Pallottini} {et~al.}(2017{\natexlab{b}}){Pallottini}, {Ferrara},
  {Gallerani}, {Vallini}, {Maiolino}, \& {Salvadori}}]{pallottini:2017_a}
{Pallottini}, A., {Ferrara}, A., {Gallerani}, S., {et~al.} 2017{\natexlab{b}},
  \mnras, 465, 2540, \dodoi{10.1093/mnras/stw2847}

\bibitem[{{Pallottini} {et~al.}(2019){Pallottini}, {Ferrara}, {Decataldo},
  {Gallerani}, {Vallini}, {Carniani}, {Behrens}, {Kohandel}, \&
  {Salvadori}}]{pallottini:2019}
{Pallottini}, A., {Ferrara}, A., {Decataldo}, D., {et~al.} 2019, \mnras, 487,
  1689, \dodoi{10.1093/mnras/stz1383}

\bibitem[{{Pallottini} {et~al.}(2022){Pallottini}, {Ferrara}, {Gallerani},
  {Behrens}, {Kohandel}, {Carniani}, {Vallini}, {Salvadori}, {Gelli},
  {Sommovigo}, {D'Odorico}, {Di Mascia}, \& {Pizzati}}]{Pallottini22}
{Pallottini}, A., {Ferrara}, A., {Gallerani}, S., {et~al.} 2022, \mnras, 513,
  5621, \dodoi{10.1093/mnras/stac1281}

\bibitem[{{Pontoppidan} {et~al.}(2022){Pontoppidan}, {Barrientes}, {Blome},
  {Braun}, {Brown}, {Carruthers}, {Coe}, {DePasquale}, {Espinoza}, {Marin},
  {Gordon}, {Henry}, {Hustak}, {James}, {Jenkins}, {Koekemoer}, {LaMassa},
  {Law}, {Lockwood}, {Moro-Martin}, {Mullally}, {Pagan}, {Player}, {Proffitt},
  {Pulliam}, {Ramsay}, {Ravindranath}, {Reid}, {Robberto}, {Sabbi}, {Ubeda},
  {Balogh}, {Flanagan}, {Gardner}, {Hasan}, {Meinke}, \&
  {Nota}}]{Pontoppidan22}
{Pontoppidan}, K.~M., {Barrientes}, J., {Blome}, C., {et~al.} 2022, \apjl, 936,
  L14, \dodoi{10.3847/2041-8213/ac8a4e}

\bibitem[{{Pontzen} {et~al.}(2013){Pontzen}, {Rovskar}, {Stinson}, {Woods},
  {Reed}, {Coles}, \& {Quinn}}]{pynbody}
{Pontzen}, A., {Rovskar}, R., {Stinson}, G.~S., {et~al.} 2013, {pynbody:
  Astrophysics Simulation Analysis for Python}

\bibitem[{{Popping}(2022)}]{Popping22}
{Popping}, G. 2022, arXiv e-prints, arXiv:2208.13072.
\newblock \doarXiv{2208.13072}

\bibitem[{{Rizzo} {et~al.}(2022){Rizzo}, {Kohandel}, {Pallottini}, {Zanella},
  {Ferrara}, {Vallini}, \& {Toft}}]{Rizzo22}
{Rizzo}, F., {Kohandel}, M., {Pallottini}, A., {et~al.} 2022, \aap, 667, A5,
  \dodoi{10.1051/0004-6361/202243582}

\bibitem[{{Rodighiero} {et~al.}(2022){Rodighiero}, {Bisigello}, {Iani},
  {Marasco}, {Grazian}, {Sinigaglia}, {Cassata}, \& {Gruppioni}}]{Rodighiero22}
{Rodighiero}, G., {Bisigello}, L., {Iani}, E., {et~al.} 2022, arXiv e-prints,
  arXiv:2208.02825.
\newblock \doarXiv{2208.02825}

\bibitem[{{Rosdahl} {et~al.}(2013){Rosdahl}, {Blaizot}, {Aubert}, {Stranex}, \&
  {Teyssier}}]{rosdahl:2013}
{Rosdahl}, J., {Blaizot}, J., {Aubert}, D., {Stranex}, T., \& {Teyssier}, R.
  2013, \mnras, 436, 2188, \dodoi{10.1093/mnras/stt1722}

\bibitem[{{Santini} {et~al.}(2022){Santini}, {Fontana}, {Castellano},
  {Leethochawalit}, {Trenti}, {Treu}, {Belfiori}, {Birrer}, {Bonchi}, {Merlin},
  {Mason}, {Morishita}, {Nonino}, {Paris}, {Polenta}, {Rosati}, {Yang},
  {Bradac}, {Calabr{\`o}}, {Dressler}, {Glazebrook}, {Marchesini}, {Mascia},
  {Nanayakkara}, {Pentericci}, {Roberts-Borsani}, {Scarlata}, {Vulcani}, \&
  {Wang}}]{Santini22}
{Santini}, P., {Fontana}, A., {Castellano}, M., {et~al.} 2022, arXiv e-prints,
  arXiv:2207.11379.
\newblock \doarXiv{2207.11379}

\bibitem[{{Sommovigo} {et~al.}(2020){Sommovigo}, {Ferrara}, {Pallottini},
  {Carniani}, {Gallerani}, \& {Decataldo}}]{Sommovigo20}
{Sommovigo}, L., {Ferrara}, A., {Pallottini}, A., {et~al.} 2020, \mnras, 497,
  956, \dodoi{10.1093/mnras/staa1959}

\bibitem[{{Teyssier}(2002)}]{teyssier:2002}
{Teyssier}, R. 2002, \aap, 385, 337, \dodoi{10.1051/0004-6361:20011817}

\bibitem[{{Topping} {et~al.}(2022){Topping}, {Stark}, {Endsley}, {Plat},
  {Whitler}, {Chen}, \& {Charlot}}]{Topping22}
{Topping}, M.~W., {Stark}, D.~P., {Endsley}, R., {et~al.} 2022, arXiv e-prints,
  arXiv:2208.01610.
\newblock \doarXiv{2208.01610}

\bibitem[{{Treu} {et~al.}(2022){Treu}, {Roberts-Borsani}, {Bradac}, {Brammer},
  {Fontana}, {Henry}, {Mason}, {Morishita}, {Pentericci}, {Wang}, {Acebron},
  {Bagley}, {Bergamini}, {Belfiori}, {Bonchi}, {Boyett}, {Boutsia},
  {Calabr{\'o}}, {Caminha}, {Castellano}, {Dressler}, {Glazebrook}, {Grillo},
  {Jacobs}, {Jones}, {Kelly}, {Leethochawalit}, {Malkan}, {Marchesini},
  {Mascia}, {Mercurio}, {Merlin}, {Nanayakkara}, {Nonino}, {Paris},
  {Poggianti}, {Rosati}, {Santini}, {Scarlata}, {Shipley}, {Strait}, {Trenti},
  {Tubthong}, {Vanzella}, {Vulcani}, \& {Yang}}]{Treu22}
{Treu}, T., {Roberts-Borsani}, G., {Bradac}, M., {et~al.} 2022, \apj, 935, 110,
  \dodoi{10.3847/1538-4357/ac8158}

\bibitem[{{Vallini} {et~al.}(2017){Vallini}, {Ferrara}, {Pallottini}, \&
  {Gallerani}}]{Vallini17}
{Vallini}, L., {Ferrara}, A., {Pallottini}, A., \& {Gallerani}, S. 2017,
  \mnras, 467, 1300, \dodoi{10.1093/mnras/stx180}

\bibitem[{{Vallini} {et~al.}(2018){Vallini}, {Pallottini}, {Ferrara},
  {Gallerani}, {Sobacchi}, \& {Behrens}}]{Vallini18}
{Vallini}, L., {Pallottini}, A., {Ferrara}, A., {et~al.} 2018, \mnras, 473,
  271, \dodoi{10.1093/mnras/stx2376}

\bibitem[{van~der Walt {et~al.}(2011)van~der Walt, Colbert, \&
  Varoquaux}]{numpy}
van~der Walt, S., Colbert, S.~C., \& Varoquaux, G. 2011, Computing in Science
  Engineering, 13, 22, \dodoi{10.1109/MCSE.2011.37}

\bibitem[{Van~Rossum \& de~Boer(1991)}]{python2}
Van~Rossum, G., \& de~Boer, J. 1991, CWI Quarterly, 4, 283

\bibitem[{Van~Rossum \& Drake(2009)}]{python3}
Van~Rossum, G., \& Drake, F.~L. 2009, Python 3 Reference Manual (Scotts Valley,
  CA: CreateSpace)

\bibitem[{{Virtanen} {et~al.}(2020){Virtanen}, {Gommers}, {Oliphant},
  {Haberland}, {Reddy}, {Cournapeau}, {Burovski}, {Peterson}, {Weckesser},
  {Bright}, {van der Walt}, {Brett}, {Wilson}, {Millman}, {Mayorov}, {Nelson},
  {Jones}, {Kern}, {Larson}, {Carey}, {Polat}, {Feng}, {Moore}, {VanderPlas},
  {Laxalde}, {Perktold}, {Cimrman}, {Henriksen}, {Quintero}, {Harris},
  {Archibald}, {Ribeiro}, {Pedregosa}, {van Mulbregt}, \& {SciPy 1. 0
  Contributors}}]{scipy}
{Virtanen}, P., {Gommers}, R., {Oliphant}, T.~E., {et~al.} 2020, Nature
  Methods, 17, 261, \dodoi{10.1038/s41592-019-0686-2}

\bibitem[{{Whitler} {et~al.}(2022){Whitler}, {Endsley}, {Stark}, {Topping},
  {Chen}, \& {Charlot}}]{Whitler22}
{Whitler}, L., {Endsley}, R., {Stark}, D.~P., {et~al.} 2022, arXiv e-prints,
  arXiv:2208.01599.
\newblock \doarXiv{2208.01599}

\bibitem[{{Witstok} {et~al.}(2022){Witstok}, {Smit}, {Maiolino}, {Kumari},
  {Aravena}, {Boogaard}, {Bouwens}, {Carniani}, {Hodge}, {Jones}, {Stefanon},
  {van der Werf}, \& {Schouws}}]{Witstok22}
{Witstok}, J., {Smit}, R., {Maiolino}, R., {et~al.} 2022, \mnras, 515, 1751,
  \dodoi{10.1093/mnras/stac1905}

\bibitem[{{Yan} {et~al.}(2022){Yan}, {Ma}, {Ling}, {Cheng}, {Huang}, \&
  {Zitrin}}]{Yan22}
{Yan}, H., {Ma}, Z., {Ling}, C., {et~al.} 2022, arXiv e-prints,
  arXiv:2207.11558.
\newblock \doarXiv{2207.11558}

\bibitem[{{Yoon} {et~al.}(2022){Yoon}, {Carilli}, {Fujimoto}, {Castellano},
  {Merlin}, {Santini}, {Yun}, {Murphy}, {Jung}, {Casey}, {Finkelstein},
  {Papovich}, {Fontana}, {Treu}, \& {Letai}}]{Yoon22}
{Yoon}, I., {Carilli}, C.~L., {Fujimoto}, S., {et~al.} 2022, arXiv e-prints,
  arXiv:2210.08413.
\newblock \doarXiv{2210.08413}

\bibitem[{{Zanella} {et~al.}(2021){Zanella}, {Pallottini}, {Ferrara},
  {Gallerani}, {Carniani}, {Kohandel}, \& {Behrens}}]{Zanella21}
{Zanella}, A., {Pallottini}, A., {Ferrara}, A., {et~al.} 2021, \mnras, 500,
  118, \dodoi{10.1093/mnras/staa2776}

\end{thebibliography}



\end{document}